\documentstyle[aps,pre,epsfig,floats]{revtex}

\begin{document}
\draft \twocolumn[\hsize\textwidth\columnwidth\hsize\csname
@twocolumnfalse\endcsname
\title{Three and four current reversals vs temperature in
correlation ratchets\\ with a simple
sawtooth potential}
\author
{Risto Tammelo*$^1$, Romi Mankin$^2$, and Dmitri Martila$^1$}
\address{
$^1$Institute of Theoretical Physics, Tartu University, 4 T\"ahe
Street, 51010 Tartu, Estonia\\
$^2$Department of Natural Sciences, Tallinn Pedagogical
University, 25 Narva Road, 10120 Tallinn, Estonia}
\date{\today}
\maketitle

\begin{abstract}
Transport of Brownian particles on a simple sawtooth potential
subjected to both unbiased thermal and nonequilibrium symmetric
three-level Markovian noise is considered. The new effects of
{\it three} and {\it four} current reversals as a function of
{\it temperature} are established in such correlation ratchets.
The parameter space coordinates of the fixed points associated
with these current reversals and the necessary and sufficient
conditions for the existence of the novel current reversals are
found.
\end{abstract}
 ]

The aim of this paper is to study current reversals in controlled
transport of Brownian particles \cite{reimann} induced by {\it
symmetric} nonequilibrium noise in ratchets with a {\it simple}
sawtooth potential, bearing in mind potential applications for
separation of nanoobjects \cite{rousselet}. It is known that
current reversals in ratchet systems can be engendered by changing
various system parameters
\cite{mielke,hondou,doering2,bier,elston,mankin1,mankin2,bartussek,kula,millonas,reimann2,derenyi,marchesoni},
including the flatness parameter of the noise
\cite{doering2,bier,elston,mankin1,mankin2}, the correlation time
of nonequilibrium fluctuations \cite{bartussek}, the temperature
in multinoise cases \cite{kula}, the power spectrum of the noise
source \cite{millonas}, the shape of the potential
\cite{reimann2}, the number of interacting particles per unit cell
\cite{derenyi}, and the mass of the particles \cite{marchesoni}.
As a rule, these results have been obtained either at the limits
of slow and fast noises or by numerical methods. At the same time,
analytic results would greatly facilitate the study, especially in
the intermediate regimes of the system parameters which is the
realm of biology. It is especially difficult to obtain analytic
expressions for the multinoise ratchets. A fortunate exception
here is the symmetric three-level telegraph process (trichotomous
noise) which is rich enough physically and can at the same time be
treated analytically \cite{mankin1,mankin2,mankin}. In this paper
we will prove, on the basis of the leading order term of a series
expansion for the probability current in terms of inverse
flatness, that there exist {\it three} and {\it four} current
reversals of the probability current as a function of {\it
temperature} $D$. Never before have more than two current
reversals with $D$ been reported for correlation ratchets with a
{\it simple} sawtooth potential. (At the same time, in the case
of rocking ratchets, infinitely many current reversals may occur
\cite{jung}.) We will also derive the necessary, and the
necessary and sufficient conditions for the existence of these
novel current reversals.
\begin{figure} \centerline{
\psfig{file=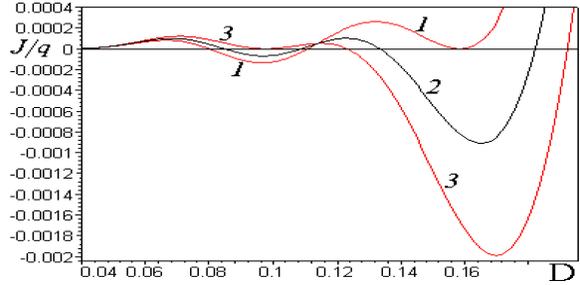,width=84.6mm,height=37.26mm} } \caption{ Four
current reversals vs temperature $D$ for fixed $d=0.0045$ and
$a=27.75$. The switching rates are respectively (1)
$\nu_1=649.651630$, (2) $\nu_2=646.770000$, and (3)
$\nu_3=643.816291$. Curve (2) has four single zeros. Curves (1)
and (3) have two single zeros and one two-fold zero. (At the fixed
values of $d$ and $a$, for curves with $\nu$ values smaller than
$\nu_3$ and greater than $\nu_1$ there can occur at most two
current reversals.) }
\end{figure}

A zero-mean trichotomous Markovian stochastic process $Z(t)$
consists of jumps between three values $z=\{a, 0, -a\},\, a\!>0$.
The jumps follow in time the pattern of a Poisson process, the
values occuring with the stationary probabilities
$P_s(a)=P_s(-a)=q$ and $P_s(0)=(1-2q)$, where $0\!<\!q\!<\!1/2$.
Denoting the state space and the transition matrix of our
trichotomous process, respectively, by $\{a_i\}\!:=\!\{a,\, 0,
-a\}$ and $T_{ij}:=P\{a_i,t+\!\tau\,|a_j,t\}$, $\; i,j=1,2,3, \;
\tau>0$, we have
\begin{equation}
\Big(T_{ij}\Big)=\Big(\delta_{ij}\Big)+\pmatrix{q-1&q&q\cr1-2q&-2q&1-2q\cr
q&q&q-1\cr}\big(1-e^{-\nu\tau}\big),\label{transmatrix}
\end{equation}
where $\nu\!>\!0$. In a stationary state, the fluctuation process
satisfies $\langle{Z(t)}\rangle=0$ and
$\langle{Z(t+\tau)Z(t)}\rangle=2qa^2\exp(-\nu\tau)$, where the
switching rate $\nu$ is the reciprocal of the noise correlation
time $\tau_c=1/\nu$, i.e. $Z(t)$ is a symmetric zero-mean
exponentially correlated noise. The trichotomous process is a
particular case of the kangaroo process \cite{doering2} with
flatness parameter $\varphi:=\langle{Z^4(t)}\rangle/
\langle{Z^2(t)}\rangle^2=1/(2q)$.

At great flatness, $q \to 0$, which is the case addressed in the
present paper, the transition matrix of our trichotomous process
takes the following form
\begin{equation}
\Big(T_{ij}\Big)=\Big(\delta_{ij}\Big)+\pmatrix{-1&0&0\cr1&0&1\cr
0&0&-1\cr}\big(1-e^{-\nu\tau}\big) + {\cal O}
(q).\label{transmatrix2}
\end{equation}
Within the framework of the three-level noise models used in
Refs.~\cite{bier,elston} flatness is determined by a parameter
$\lambda$, which regulates the relative amount of time spent in
the state $z=0$ as opposed to the states $z=a$ and $z=-a.\;$ If
$\lambda \to 0$, the flatness $\varphi \to \infty$ and the
leading order terms in the transitions matrixes of the noise
processes of Refs.~\cite{bier,elston} and Eq.~(\ref{transmatrix2})
become equivalent. Thus, at great flatness our trichotomous noise
practically coincides with the noise used by Bier \cite{bier} and
Elston-Doering \cite{elston}.

\begin{figure}
\centerline{ \psfig{file=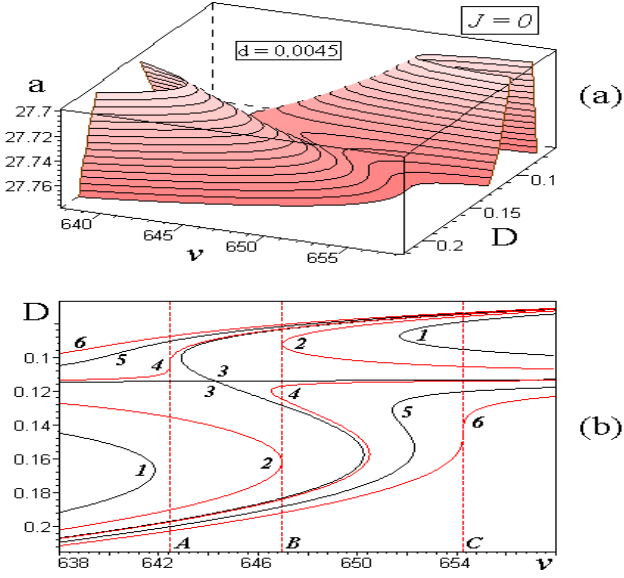,width=84.6mm,height=76.44mm} }
\caption{(a) The surface of current reversals $J(D,a,\nu)=0$ for a
fixed asymmetry parameter $d=0.0045$. (b) The projection of the
surface onto plane $(D,\nu)$. The level curves correspond to the
following values of the noise amplitude: (1)~$a=27.700000$,
(2)~$a=27.733329$, (3)~$a=27.753662$, (4)~$a=27.755045$,
(5)~$a=27.765482$, (6)~$a=27.775920.$ The effect begins at
$\nu_{A}=642.4480$ and ends at $\nu_{C}=654.2591$; whereas at
$\nu_{B}=646.9504$ there occur two two-fold zeros.}
\end{figure}

We describe overdamped motion of Brownian particles in
dimensionless units by the Langevin equation
\begin{equation}
{dX\over dt}= h(X)+\xi(t)+Z(t),\quad h(x)\equiv-{dV(x)\over dx},
\label{langevin}
\end{equation}
where $V(x)=V(x+1)$ is a periodic spatial potential of period 1.
The thermal noise satisfies $\langle\xi(t)\rangle=0$ and
$\langle\xi(t_1)\xi(t_2)\rangle$ $=2D\delta(t_1 -t_2)$, where $D$
is the thermal noise strength which will below simply be called
the temperature, for the sake of brevity. As said, we take the
random force $Z(t)$ to be a zero-mean trichotomous Markovian
stochastic process \cite{mankin1,mankin2,mankin}. To derive an
exact formula for $J$, the Fokker-Planck master equation
corresponding to Eq.~(\ref{langevin}) is used, supposing that the
potential $V(x)$ in Eq.~(\ref{langevin}) is piecewise linear
(sawtooth-like) and its asymmetry is determined by a parameter
$d\in(0,1)$, with $d=1/2$ for symmetric $V(x)$. The force caused
by the potential is $ h(x)=b:=1/d $ for $ x\in(0,d) $ and $
h(x)=-c:=-1/(1-d) $ for $ x\in(d,1) $. Under these assumptions, a
complex exact formula as a quotient of two eleventh-order
determinants can be derived for the probability current $J$. To
obtain a more manageable formula, on the assumption that the
flatness parameter is large, $\varphi=1/(2q)\gg 1$, we can expand
the current in a series $J=qJ^{(1)}+q^2J^{(2)}+\ldots\;$ One of
the present authors derived for the leading order term $qJ^{(1)}$
a rather lengthy analytic expression, i.e.,  Eqs.~(31)-(33) in
Ref.~\cite{mankin2}. Herein we shall apply graphical analysis to
study the analytic expression of $qJ^{(1)}$ in the intermediate
regimes.

\begin{figure}
\centerline{ \psfig{file=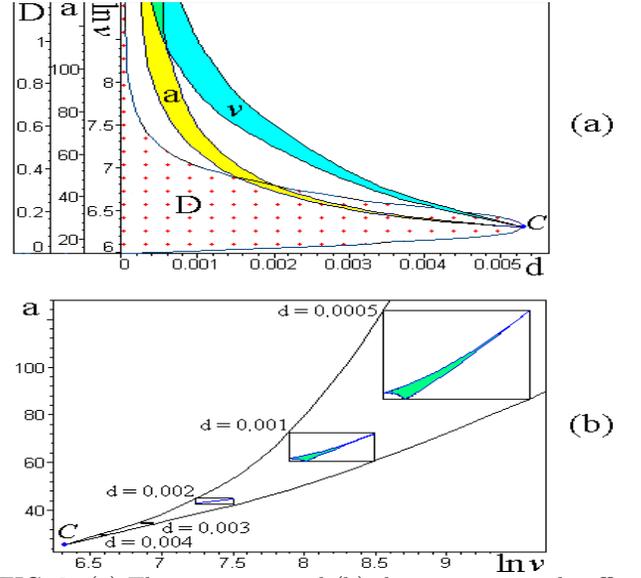,width=84.6mm,height=76.03mm}}
\caption{(a) The necessary and (b) the necessary and sufficient
conditions for the four-current-reversal effect. The dotted region
in (a) displays the possible range of the four zeros of the
current $J(D)$ at different values of $d$.}
\end{figure}

First we will briefly review the asymptotic limits of the current
$J$ as the function of $\nu$ and $D$, found in
Ref.~\cite{mankin2}. In the case of large flatness at the
asymptotic limits of both small and large $\nu$ the function
$J=J(\nu)$ is always positive. Hence, in the case under
discussion there can exist either none or an even number of
current reversals with $\nu$. Two current reversals vs $\nu$ (as
well as $D$) were addressed by Ref.~\cite{mankin2}. The case of
four current reversals vs $\nu$ with the necessary and sufficient
conditions for their existence was extensively discussed in our
earlier paper \cite{mankin3}.

What concerns current reversals as functions of $D$, then in the
asymptotic limit of high temperature, $D\to\infty$, we find that
\begin{equation}J\approx{q(b-c)a^2\over 180bcD^4}.
\label{jhigh}
\end{equation}
Thus, at high temperatures the behavior of the function $J=J(D)$
is uniform: the current is always positive and decreases
monotonically to zero as $D \to \infty$. At low temperatures, in
the cases $a<c$ and $c<a<b$ the current is positive for all
values of $\nu$ and $d$. For $a>b$ the current behaves
asymptotically as
\begin{eqnarray}
J\approx q\nu
\left\{\left[e^{\nu/c(a-c)}-e^{-\nu/b(a+b)}\right]^{-1}\right.-\nonumber\\
-\left.\left[e^{\nu/b(a-b)}-e^{-\nu/c(a+c)}\right]^{-1}\right\}.
\label{jlow}
\end{eqnarray}

\begin{figure}
\centerline{ \psfig{file=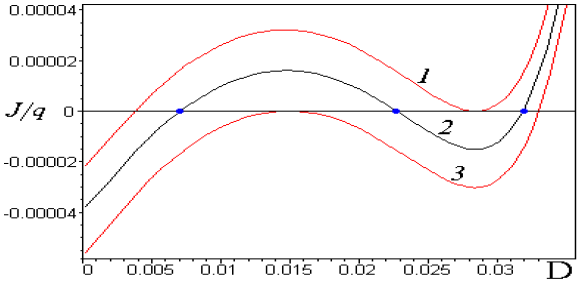,width=84.6mm,height=37.26mm} }
\caption{Three current reversals vs temperature $D$ for fixed
$d=0.005$ and $a=255.86$. Switching rates are
(1)~$\nu_1=3.17379365$, (2)~$\nu_2=3.17381112$, and
(3)~$\nu_3=3.17382860$. Curve (2) has three single zeros. Curves
(1) and (3) have one single and one two-fold zero.}
\end{figure}

Notably, $J$ is positive in the case of $a\le b c$ at any $\nu$.
If $a>bc$, then the current reverses to negative at a critical
value $\nu=\nu_0$. The point of reversal $\nu_0$, being a
nontrivial solution of the transcendental equation $J(\nu_0)=0$,
can be found by numerical calculation from Eq.~(\ref{jlow}). In
the limit of low temperatures, $D \to 0$, the behavior of the
function $J=J(D)$ is {\it not} uniform. Depending on the values of
the remaining parameters of the system, at $D=0$ the function
$J=J(D)$ may have a (finite) positive, a (finite) negative or a
zero value and may start to increase, decrease or remain
practically constant for a while as the temperature $D$ increases
\cite{mankin2}. What is important about the asymptotic limits in
the context of the present paper, where we are interested in the
behavior of the current in the intermediate domains of the system
parameters, is the fact that at the limit of large $D$ the
function $J=J(D)$ is always positive, whereas at the limit of
small $D$ it can be either negative or positive, and,
consequently, there can occur {\it any} number, odd or even, of
current reversals with $D$.

Before the present paper not more than two current reversals with
respect to $D$ had been found for correlation ratchets with a
simple sawtooth potential. Moreover, in a paper \cite{mankin2} by
one of the authors it was even argued that the possible number of
current reversals with temperature is either zero, one, or two.

Next we will examine the four-current-reversal effect as a
function of $D$ (see Fig.~1). Fig.~2(b) exhibits the level curves
of zero current, $J(D,\nu;d,a)=0$, for fixed $d=0.0045$ at
different fixed values of $a=\text{const}$. The level curves may
be considered as functions $\nu=\nu(D)$ (as well as $D=D(\nu)$),
with $d$ and $a$ being parameters. In Fig.~2(b) the level curves
on the left  close at smaller finite values of $\nu$ (not shown),
whereas both branches of the level curves on the right approach
zero as $\nu$ grows. Regarding the upper branch on the right, if
$a <d^{-1}(1-d)^{-1}$, then $D$ becomes zero only at the limit
$\nu \to\infty$, whereas if $a > d^{-1}(1-d)^{-1}$, then $D$ is
zero at a finite $\nu$. In view of this, two types of the level
curves are distinguishable in Fig.~2(b): namely, the connected
ones (i.e., curves (4)-(6)) and the ones (i.e., curves (1) and (2)
comprising two components, viz., a closed curve and a curve with
one end open. There is one very special level curve (i.e., curve
(3)) which intersects itself at the saddle point. The four current
reversals vs $D$ effect exists at a certain fixed value of
$\nu_{\rm{fixed}}=\text{const}$ if and only if there exist $d$ and
$a$ for which the function $\nu=\nu(D;d,a)$ has a local maximum
$\nu_{\text{max}}> \nu_{\rm{fixed}}$ and a local minimum
$\nu_{\text{min}} < \nu_{\rm{fixed}}$ (see the vertical dashed
lines in Fig.~2(b)).

By gradually varying $a$ and $d$ one can demonstrate that the
region of existence of the four-current-reversal effect as a
function of $D$ shrinks to a $4$-point $C$, which has the
following coordinates: $d_C\approx0.0053305, a_C\approx
25.3379050, D_C\approx0.1244425$ and $\ln\nu_C\approx6.2977305$.
The four-current-reversal effect is possible if $d\in (0,d_C),
a\in(a_C,\infty)$ and $\nu\in (\nu_C,\infty)$ (see Fig.~3). The
necessary conditions for the existence of the
four-current-reversal effect are shown in Fig.~3(a) by the shaded
regions in planes $(d,a)$ and $(d,\nu)$, and in Fig.~3(b) by the
region between the ascending outermost curves in plane
$(\nu,a)$,  while the intensively shadowed narrow wedge-shaped
areas in Fig.~3(b) fix the values of $d,a,$ and $\nu$, which are
necessary and sufficient for the existence of the
four-current-reversal effect as a function of $D$.

\begin{figure}
\centerline{\psfig{file=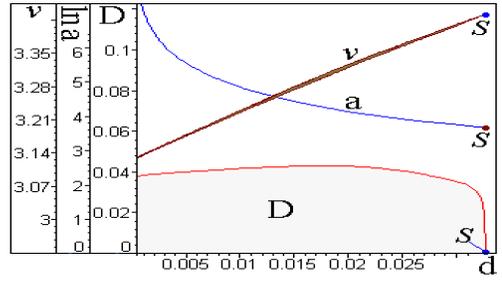,width=84.6mm,height=37.26mm}}
\caption{The necessary conditions for the
three-\-current\-reversal vs $D$ effect. The curves $a=a(d)$ and
$\nu=\nu(d)$ represent actually narrow regions in the planes
$(d,a)$ and $(d,\nu)$. Especially narrow is the bow depicted by
$a=a(d)$, e.g., at $d=0.001$ the noise ranges from
$a_{\rm{min}}=1280.330$ to $a_{\rm{max}}=1280.462$.}
\end{figure}

What concerns the three-current-reversal effect as a function of
temperature $D$ (see Fig.~4), then the effect occurs only in a
very narrow range of the system parameters $a$ and $\nu$ (see
Fig.~5). By varying $a$ and $d$ step by step, we will see that
the region of existence of the three-current-reversal effect as a
function of $D$ shrinks to a critical $4$-point $S$ in the
parameter space, which has the following coordinates:
\(d_{S}\approx0.03275\), \(D_{S}=0\), \(a_{S}\approx38.95953\),
\(\nu_{S}\approx3.42833\). The three-current-reversal effect is
possible if $d\in (0,d_S)$, $\nu\in (3.1275,\nu_S)$ and
$a\in(a_S,\infty)$ (see Fig.~5).

As demonstrated above, the necessary and sufficient conditions for
the existence of an odd number of  current reversals vs $D$ are $a
> bc$ and $\nu > \nu_0$, where $\nu_0$ is a zero of the function $J
(\nu)$, see Eq.~(\ref{jlow}). In the opposite case, the number of
current reversals is even or zero. To elucidate the physical
meaning of the above conditions, let us re-derive
Eq.~(\ref{jlow}) on the grounds of the following physical
considerations. In the case of great flatness, $q \ll 1$, the
noise $Z(t)$ is with overwhelming probability, $P_s(0)=1-2q
\approx 1$, at the state $z=0$ and the current $J$ can be
regarded as the sum $J=J^+ + J^-$, where the positive current
$J^+$ is caused by the transitions $z\!=\!a \leftrightarrow
z\!=\!0$ and the negative current $J^-$ by the transitions
$z\!=\!-a \leftrightarrow z\!=\!0$. (Note that the transitions
$z\!=\!a \leftrightarrow z\!=\!-a$ induce current which is
proportional to $q^2$ and will be discarded at the present
approximation.) Under these assumptions the stationary
probability distribution at the noise state $z=0$ with $D=0$
consists, evidently, of delta functions at $x= \pm n+d,\; n=
0,1,2, \dots\;$ Within the interval $(0,1)$ the center of mass
lies at $y_0 =d$. Let at the initial moment occur the transition
$z\!=\!0 \rightarrow z\!=\!-a$. The first time when the noise
turns back to $z=0$ is denoted by $t_0$. The center of mass
stabilizes now at the position $y$. It is easy to find that the
center of mass is shifted by $\Delta y(t_0) =y-y_0$ with $\Delta
y (t_0) = -(n+1)$ if $(n+1) T^- > t_0 \geq n T^- + T_1^-$ and
$\Delta y(t_0) = -n$ if $n T^- + T_1^-
> t_0 \geq n T^-$. The interval $T^-=T_1^- + ((1-d) /(a +c))$ is
the time that the particle for $z=-a$ takes to pass the period
$L=1$ of the potential; $T_1^- = d/(a-b)$ is the time necessary
for passing the length $d$ (i.e., from the minimum of the
potential to the maximum in the negative direction). The
probability $W(t_0)$ that in a certain time interval $(0,t_0)$ the
transitions $z\!=\!-a \rightarrow z\!=\!0$ do not occur, is given
by $W (t_0) = \exp\, (-\nu t_0)$. The probability that such a
transition will occur within the time interval $(t_0, t_0 + dt_0)$
is $\nu dt_0$, and consequently, $<\Delta y> = \nu \int_0^{\infty}
e^{-\nu t_0} \Delta y (t_0) dt_0$. Considering that the average
number of transitions per unit time into the $z\!=\!-a_0$ state
is $q \nu$, we obtain $J^-\!= q \nu <\Delta y>\, =\! -q \nu W
(T_1^-) /[1-W(T^-)]$. Similarly, one can derive the positive
component of the current, viz. $J^+\!= q \nu W (T_1^+) / [1-W
(T^+)]$, where $T_1^+\!= (1-d)/(a-c)$, $T^+\!= T_1^+\!+ d/(a+b)$.
The inequality $T^+\!< T^-$ being equivalent to $d<1/2$ , it is
evident that the total current $J = J^+\!+ J^-$, whose expression
coincides exactly with Eq.~(\ref{jlow}), can be negative only
when $T_1^-\!< T_1^+$. The latter inequality can be written as
$a>bc$, which is just the necessary condition.

Following similar trains of thought as above in the case of low
temperature $D \ll 1$ and replacing the passage times $T^\pm$ and
$T_1^\pm$ by respective mean first passage times  $<T^\pm>$ and
$<T_1^\pm>$, one can re-derive the necessary conditions for the
existence of the three-current-reversal effect as a function of
temperature. Namely, we obtain the formulas $a=41/(32d)\!-\!0.25$
and $\nu_0\!=3.12(1\!+3d)$ which quite well approximate the
corresponding curves in Fig.~5.

Finally we will discuss the possible usefulness of the phenomena
of three and four current reversals. One of the first applications
of the Brownian ratchet mechanism has been to separate particles
by size, charge, mass, etc. Separation can be achieved even if
there are no current reversals, as particles of different viscous
friction move at different speeds. If there is only one current
reversal, then particles of different friction coefficients. move
in the opposite directions within the same environment. If there
occur two current reversals with $\nu$, then particles with
parameter values within a characteristic interval can be
separated, as they move in the direction opposite to that of the
remaining particles. However, it is known that the two zeros of
the current $J(\nu)$ generally occur at largely displaced values
of $\nu$. Therefore, in the opposite direction move those
particles whose friction coefficients are within a wide interval
and the separation effect is of low selectivity. Though this wide
interval can be made narrower by varying the other system
parameters, the really narrow intervals occur only in the
vicinity of the transition regimes (i.e., the transition from two
current reversals to zero current reversals), where the absolute
value of the current is small. On the other hand, the
three-current-reversal with $D$ effect enables us to design
continuous two-step separation schemes with very high selectivity
within a narrow $\nu$ interval. Note that at a fixed value of $d$
the range of values of $\nu$ where the three-current-reversal vs
$D$ effect exists is very narrow indeed (see Fig.~5).

Within the framework of the calculation scheme of the present
paper, the absolute value of the net current is inversely
proportional to the flatness parameter of the trichotomous noise,
$\varphi=1/(2q)$, and $q$ as an expansion parameter is generally
considered to be infinitesimal. However, we have managed to show
by direct numerical calculations without using expansion in $q$
that the effects are present up to $q\approx 0.015$.

Beyond the separation methods, the phenomena of three and four
current reversals may be of interest in biology, e.g., when
considering the motion of macromolecules. While it is known that
the two-current-reversal effect allows one pair of motor-proteins
to move simultaneously in opposite directions along the
microtubulae inside the eucariotic cells, then existence  of
three and four current reversals will enable such a simultaneous
motion of many pairs of motor proteins. Whether the intracellular
transport makes use of the flashing ratchet mechanism as one
presently tends to think, or the correlation ratchets mechanism,
or a combination of them remains to be seen.

To summerize, it is remarkable that the interplay of the {\it
symmetric} three-level and thermal noises in the ratchets with a
{\it simple} sawtooth potential generates such a rich variety of
cooperation effects as up to four current reversals with
temperature as well as switching rate. The results are the more
surprising because in analogous model systems with the {\it
symmetric} dichotomous noise the current reversals are altogether
absent.

We acknowledge partial support by the Estonian Science Foundation
Grants Nos 4042 and 4208.

\end{document}